\documentclass{amsart}[10 point]

\usepackage{pstricks,pst-node,pst-plot}
\usepackage{epsfig}
\usepackage{latexsym}
\usepackage{amsmath}
\usepackage{mathrsfs}
\newtheorem{theorem}{Theorem}[section]
\newtheorem{proposition}[theorem]{Proposition}

\newcommand{\RR}{{\mathbb R}}
\newcommand{\PP}{{\mathbb P}}
\newcommand{\EE}{{\mathbb E}}
\newcommand{\V}{\mathop{\textrm{Var}}\nolimits}
\newcommand{\Cov}{\mathop{\textrm{Cov}}\nolimits}

\newcommand{\A}{{\mathscr A}}
\newcommand{\B}{{\mathscr B}}

\newcommand{\LL}{{\mathscr L}}
   
\newcommand{\old}[1]{{}}

\title[Trait-dependent extinction]{Trait-dependent extinction leads to greater expected biodiversity loss}
\author{Be\' ata Faller and Mike Steel}

\thanks{We thank the New Zealand Marsden Fund and the Allan Wilson Centre for Molecular Ecology and Evolution for supporting this work, and Stefan Gr{\"u}newald for useful discussions on a related problem}

\address{Allan Wilson Centre for Molecular Ecology and Evolution, Department of Mathematics and
  Statistics, University of Canterbury, Christchurch, New Zealand}

\email{fallerbeata@gmail.com, m.steel@math.canterbury.ac.nz}

\subjclass{05C05; 92D15}

\keywords{tree, Markov process, AD inequality, FKG inequality, lattice, phylogenetic
diversity}

\begin{document}
 \begin{abstract}
We use a classical combinatorial inequality to establish a Markov inequality for multivariate binary Markov processes on trees. We then apply this result, alongside with the FKG inequality, to compare the expected loss of biodiversity under two  models of species extinction. One of these models is the generalized version of an earlier model in which extinction is influenced by some trait that can be classified into two states and which evolves on a tree according to a Markov process. Since more than one trait can affect the rates of species extinction, it is reasonable to allow, in the generalized model, $k$ binary states that influence extinction rates. We compare this model to one that has matching marginal extinction probabilities for each species but for which the species extinction events are stochastically independent.
\end{abstract}
\maketitle

\newpage

\section{Introduction}

The concept of a `Markov process on a tree' generalizes the notion of a Markov chain and has been extensively studied in physics, information theory, and evolutionary biology. In evolution it is used to model the stochastic evolution of traits on a phylogenetic tree \cite{fel, sem}. In this work, we establish a generic Markov inequality for multivariate Markov processes that consist of $k$ independent but not necessarily identical two-state Markov processes on a tree. The inequality has been specifically designed for the purpose of comparing a new species extinction model with existing ones in conservation biology. This new model is the generalized version of the `s-FOB' model \cite{stf}, in which the extinction risk of a species is associated with an underlying state that evolves on an evolutionary tree. In the more general setting, extinction is influenced by $k$ independently evolved traits rather than only one, giving a more realistic model.

We compare the expected loss and the variance of `phylogenetic diversity' under this model to the corresponding values of a simpler model in which extinction events are treated independently.
We show that when extinction events reflect the evolutionary history of many characteristics,
the expected loss of phylogenetic diversity is greater than
or equal to that predicted under a model with independent extinction events. This generalizes the result presented in Section 3 of \cite{stf}, and suggests that simple models that treat species extinctions independently may systematically underestimate the loss of phylogenetic diversity.

Given this inequality between the expected future phylogenetic diversity under these two models we might expect a similar inequality to apply for the variance. However, we show that there is no similar relationship between the variances corresponding to the two models. There are examples for which the variance of future phylogenetic diversity under an independent extinction scenario can be either smaller or greater than the variance under the model in which extinction events are influenced by $k$ characteristics, even for $k=1$.

In the next section, we define the multivariate Markov processes under scrutiny and then state and prove the Markov inequality. To demonstrate the phylogenetic application, Section~\ref{applic} presents the inequality between the expected loss of phylogenetic diversity and our findings concerning the variance of future phylogenetic diversity.

\section{The Markov inequality}
\label{sectwo}

Let $T$ be a rooted tree with root vertex $\rho$ and with leaf set $X$. Consider $k$ independent, non-identical two-state Markov processes on $T$, each of which with the state space $\{0,1\}$ (for a formal definition of Markov processes on trees, see, for example, \cite{cha, sem, ste}).
For each vertex
$v$ of $T$ and for $j=1,\ldots,k$, let $\xi_j(v)$ denote the random state that $v$ is assigned in the $j$th Markov process. Furthermore, for $j=1,\ldots,k$ and for $i\in\{0,1\}$, let $\pi_i^{(j)}$ be the probability that $\xi_j(\rho)=i$.
Viewing the edges of $T$ as arcs directed away from the root, let $P^{(j)}(r,s)$ be the transition matrix assigned to arc $(r,s)$ in the $j$th process. The $il$-entry $P^{(j)}(r,s)_{il}$ of this $2 \times 2$ matrix is, by definition, the conditional probability that $\xi_j(s)=l$ given that $\xi_j(r)=i$.
For each $j$, having specified the probabilities $\pi_i^{(j)}$ and the transition matrices $P^{(j)}(r,s)$, $i\in\{0,1\},(r,s)\in A_T$ (the arc set of $T$), the $j$th Markov process on $T$ is uniquely defined \cite{cha, sem, ste}.

We now combine these $k$ Markov processes into a vector  (having $j$th coordinate $\xi_j$) to provide a multivariate Markov process on $T$ with state space  $\{0,1\}^k$.  In this process, each vertex $v$ of $T$ is assigned state $\mbox{\boldmath $\xi$}(v)=(\xi_1(v),\ldots,\xi_k(v))$. Let $\mathbf{i}=(i_1,\ldots,i_k)\in\{0,1\}^k$ and let $\pi_\mathbf{i}$ be the probability that $\mbox{\boldmath $\xi$}(\rho)=\mathbf{i}$. Then, by the independence of the $k$ processes, we get $\pi_\mathbf{i}=\prod_{j=1}^k\pi_{i_j}^{(j)}$. Similarly, for the transition matrix $P(r,s)$ correspondning to arc $(r,s)$ in the multivariate process, the entry $P(r,s)_{\mathbf{il}}$ in `row $\mathbf{i}$' and `column $\mathbf{l}$' (for $\mathbf{i}=(i_1,\ldots,i_k),\mathbf{l}=(l_1,\ldots,l_k) \in \{0,1\}^k$) becomes $\prod_{j=1}^kP^{(j)}(r,s)_{i_jl_j}$. This is the conditional probability that $\mbox{\boldmath $\xi$}(s)=\mathbf{l}$ given that $\mbox{\boldmath $\xi$}(r)=\mathbf{i}$.
With these, the multivariate Markov process is uniquely defined. 

We will assume throughout that all the $\pi$ values are strictly positive and that $\det P^{(j)}(r,s) \geq 0$ for each arc $(r,s)$ and for each $j$. Note that this implies that $\det P(r,s)\geq 0$. Namely, it can be seen that $P(r,s)$ is the {\em Kronecker product} of the $k$ matrices $P^{(j)}(r,s)$, and so $\det P(r,s)=(\det P^{(1)}(r,s)\times\ldots\times \det P^{(k)}(r,s))^{2^{k-1}}$ (see \cite{kron} for the definition and properties of the Kronecker product).
However, we are neither assuming that any of the $k$ processes are identical, nor that within any of them, the arcs are assigned the same transition matrix.

Consider now a realization $\mathbf{U}=(U_1,\ldots,U_k)$ of $\mbox{\boldmath $\xi$}=(\xi_1,\ldots,\xi_k)$.
Note that $\mathbf{U}$ is a function from $V$ into the set $\{0,1\}^k$ of character states.
Let $P(\mathbf{U})$ denote the probability that $\mbox{\boldmath $\xi$}=\mathbf{U}$, that is, the probability that for each $v \in V$, $v$ is assigned $\mathbf{U}(v)$.
For $j=1,\ldots,k$, let $\delta_j(\mathbf{U},v) =0$ if the $j$th coordinate $U_j(v)$ of $\mathbf{U}(v)$ is $0$ and let $\delta_j(\mathbf{U},v) =1$ if $U_j(v)=1$.
Also, let $\delta(\mathbf{U},v)$ denote the state that $v$ is assigned in $\mathbf{U}$.
Now we are able to express $P(\mathbf{U})$ in terms of the transition matrices and the $\pi$ values of the multivariate process, using the Markov property (we follow \cite{sem}). We have:
\begin{eqnarray}
\label{product}
P(\mathbf{U}) &=& \pi_{\delta(\mathbf{U}, \rho)}\cdot \prod_{(r,s) \in A_T}P(r,s)_{\delta(\mathbf{U},r)\delta(\mathbf{U},s)},\nonumber\\
\noalign{\noindent which, by the independence of the $k$ two-state processes, gives:}\nonumber\\
P(\mathbf{U}) &=&\prod_{j=1}^k \pi_{\delta_j(\mathbf{U},\rho)}^{(j)}\cdot\prod_{(r,s) \in A_T}\prod_{j=1}^k P^{(j)}(r,s)_{\delta_j(\mathbf{U},r)\delta_j(\mathbf{U},s)}\\
&=& \prod_{j=1}^k \left(\pi_{\delta_j(\mathbf{U},\rho)}^{(j)}\prod_{(r,s) \in A_T}P^{(j)}\left(r,s\right)_{\delta_j(\mathbf{U},r)\delta_j(\mathbf{U},s)}\right).\nonumber
\end{eqnarray}

Recall that a {\em lattice} $\LL$ is a partially ordered set in which any two elements $a,b\in\LL$ have a unique least upper bound $a\vee b$, called their {\em join}, and a unique greatest lower bound $a\wedge b$, which is their {\em meet}. A lattice is {\em distributive} if $a\wedge (b\vee c)=(a\wedge b)\vee(a\wedge c)$ for all $a,b,c\in\LL$ or equivalently $a\vee (b\wedge c)=(a\vee b)\wedge(a\vee c)$ for all $a,b,c\in\LL$.

Let $\LL_V$ be the set of all possible realizations of $\mbox{\boldmath $\xi$}$. 
Let $\mathbf{Y},\mathbf{Z}\in \LL_V$, and let $\leq$ be the partial order over $\LL_V$ in which $\mathbf{Y}\leq \mathbf{Z}$ whenever $Y_j(v) \leq Z_j(v)$ for each vertex $v\in V$ and for each $j=1,\ldots,k$, and in which $\mathbf{Y}$ and $\mathbf{Z}$ are incomparable otherwise.
Clearly, any two elements $\mathbf{Y}$ and $\mathbf{Z}$ of the partially ordered set $(\LL_V,\leq)$ have a join $\mathbf{Y}\vee \mathbf{Z}$ and a meet $\mathbf{Y}\wedge \mathbf{Z}$. These are the realizations of $\mbox{\boldmath $\xi$}$ that, to each vertex $v\in V$, assign state $(\max\{Y_1(v),Z_1(v)\},\ldots,\max\{Y_k(v),Z_k(v)\})$ and state $(\min\{Y_1(v),Z_1(v)\},\ldots,\min\{Y_k(v),Z_k(v)\})$, respectively. It follows that $(\LL_V,\vee,\wedge)$ is a lattice on $\LL_V$. It is easy to see that this lattice is distributive.

Recall that $X$ denotes the leaf set of $T$ and fix a non-empty subset $W$ of $X$. For each function $\mathbf{U}$ in $\LL_V$, define $\mathbf{u}=(u_1,\ldots,u_k)$ to be the restriction of $\mathbf{U}$ to $W$; that is, $\mathbf{u}=\mathbf{U}|_W$. With this we have $\mathbf{u}(v)=\mathbf{U}(v)$ for each leaf $v$ in $W$. Since $\mathbf{u}$ is a function from the non-empty subset $W$ of $X$ into a set of character states, it is also called a {\em character on} $X$ \cite{sem}.
Let $\LL_W$ be the set that contains, for each $\mathbf{U}\in \LL_V$, the restricted function $\mathbf{u}=\mathbf{U}|_W$.
Let $\mathbf{y},\mathbf{z}\in \LL_W$, and let $\leq$ be the partial order over $\LL_W$ such that if $y_j(v) \leq z_j(v)$ for each $v\in W$ and for each $j=1,\ldots,k$, we have $\mathbf{y}\leq \mathbf{z}$; otherwise $\mathbf{y}$ and $\mathbf{z}$ are incomparable.
The join $\mathbf{y}\vee \mathbf{z}$ and the meet $\mathbf{y}\wedge \mathbf{z}$ can be obtained for any two elements $\mathbf{y},\mathbf{z}$ of $\LL_W$ analogously to the case of $\LL_V$, defining the finite distributive lattice $(\LL_W,\vee,\wedge)$. Now let $p({\mathbf{u}})$ be the probability that for each leaf $v$ in $W$, $v$ is assigned $\mathbf{u}(v)$.

This marginal probability is given by:
\begin{equation}
\label{Asets}
p({\mathbf{u}})=\sum_{\mathbf{U} \in \A_{\mathbf{u}}}P(\mathbf{U}),  \mbox{  where  }  \A_\mathbf{u} := \{\mathbf{U} \in \LL_V:\mathbf{U}|_W=\mathbf{u}\}.
\end{equation}

An example to illustrate this concept is provided in Figure~1.

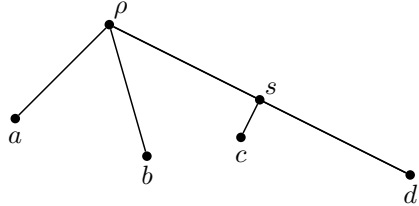
\begin{figure}[htp]
\begin{center}
\begin{pspicture}(0,0)(5.5,2.75)

\dotnode(0.25,1.25){t1}
\dotnode(1.5,2.5){t2}
\dotnode(2,0.75){t3}
\dotnode(3.5,1.5){t4}
\dotnode(5.5,0.5){t5}
\dotnode(3.25,1){t6}

\ncline[linewidth=0.02]{t1}{t2}
\ncline[linewidth=0.02]{t2}{t3}
\ncline[linewidth=0.02]{t2}{t5}
\ncline[linewidth=0.02]{t2}{t5}
\ncline[linewidth=0.02]{t4}{t6}

\rput(1.65,2.7){\rnode{m1}{$\rho$}}
\rput(3.65,1.65){\rnode{m2}{$s$}}
\rput(0.25,1){\rnode{m3}{$a$}}
\rput(2,0.5){\rnode{m4}{$b$}}
\rput(3.25,0.75){\rnode{m5}{$c$}}
\rput(5.5,0.25){\rnode{m6}{$d$}}

\end{pspicture}
\end{center}\caption{Let $k=1$ and let $\mathbf{u}$ be denoted by $u$. In this example, if $W=\{a,b,c,d\}$, $u(a)=u(c)=0$, and $u(b)=u(d)=1$, then $p(u)=\pi_0P(\rho,a)_{00}P(\rho,b)_{01}P(\rho,s)_{00}P(s,c)_{00}P(s,d)_{01}+
\pi_0P(\rho,a)_{00}P(\rho,b)_{01}P(\rho,s)_{01}P(s,c)_{10}P(s,d)_{11}+
\pi_1P(\rho,a)_{10}P(\rho,b)_{11}P(\rho,s)_{10}P(s,c)_{00}P(s,d)_{01}+
\pi_1P(\rho,a)_{10}P(\rho,b)_{11}P(\rho,s)_{11}P(s,c)_{10}P(s,d)_{11}$.}\label{oldfig}
\end{figure}

The following proposition extends a result from \cite{stf}, which dealt with the special case $k=1$.

\begin{proposition}
\label{subadd}
Consider $k$ independent two-state Markov processes on a tree with leaf set $X$. Assume that for each of them, all the determinants of the transition matrices are non-negative. Then, for the corresponding multivariate process and for any two characters $\mathbf{y},\mathbf{z}\colon W\to \{0,1\}^k$ on $X$ from a fixed non-empty subset $W$ of $X$, we have:
$$p({\mathbf{y}})\cdot p({\mathbf{z}}) \leq p({\mathbf{y}\vee\mathbf{z}})\cdot p({\mathbf{y}\wedge\mathbf{z}}).$$
\end{proposition}

\begin{proof}
Consider any two elements $\mathbf{Y}$ and $\mathbf{Z}$ of $\LL_V$. We first prove the following:
\begin{equation}
\label{ineq1}
P(\mathbf{Y})\cdot P(\mathbf{Z}) \leq P(\mathbf{Y} \vee \mathbf{Z}) \cdot P(\mathbf{Y} \wedge \mathbf{Z}).
\end{equation}

Denote the term 
in the brackets of equation (\ref{product}) by $P_j(\mathbf{U})$ to get $P(\mathbf{U})=\prod_{j=1}^k P_j(\mathbf{U})$.
Applying this to $\mathbf{U} \in \{\mathbf{Y}, \mathbf{Z}, \mathbf{Y}\vee \mathbf{Z}, \mathbf{Y}\wedge \mathbf{Z}\}$, inequality (\ref{ineq1}) can be written as $\prod_{j=1}^k P_j(\mathbf{Y}) \prod_{j=1}^k P_j(\mathbf{Z})\leq \prod_{j=1}^k P_j(\mathbf{Y}\vee \mathbf{Z}) \prod_{j=1}^k P_j(\mathbf{Y}\wedge \mathbf{Z})$. It is clear that proving $P_j(\mathbf{Y})P_j(\mathbf{Z})\leq P_j(\mathbf{Y}\vee \mathbf{Z}) P_j(\mathbf{Y}\wedge \mathbf{Z})$ for each $j$ establishes (\ref{ineq1}).

So let $j$ be an arbitrary index in $\{1,\ldots,k\}$ and consider the products $P_j(\mathbf{Y})P_j(\mathbf{Z})$ and $P_j(\mathbf{Y}\vee \mathbf{Z}) P_j(\mathbf{Y}\wedge \mathbf{Z})$. These can each be written as a product of two $\pi^{(j)}$ values multiplied by a product over the arcs $(r,s)$ of $T$ of two entries of $P^{(j)}(r,s)$. The products of the two $\pi^{(j)}$ terms agree
in $P_j(\mathbf{Y})P_j(\mathbf{Z})$ and $P_j(\mathbf{Y}\vee \mathbf{Z}) P_j(\mathbf{Y}\wedge \mathbf{Z})$, that is, $\pi_{\delta_j(\mathbf{Y},\rho)}^{(j)}\pi_{\delta_j(\mathbf{Z},\rho)}^{(j)} = \pi_{\delta_j(\mathbf{Y}\vee \mathbf{Z},\rho)}^{(j)}\pi_{\delta_j(\mathbf{Y}\wedge \mathbf{Z},\rho)}^{(j)}$. The products of the two
$P^{(j)}(r,s)$ entries agree in $P_j(\mathbf{Y})P_j(\mathbf{Z})$ and $P_j(\mathbf{Y}\vee \mathbf{Z}) P_j(\mathbf{Y}\wedge \mathbf{Z})$, except for the cases in which either (i) $\delta_j(\mathbf{Y},r)=0$, $\delta_j(\mathbf{Y},s)=1$, $\delta_j(\mathbf{Z},r)=1$ and $\delta_j(\mathbf{Z},s)=0$, or (ii) $\delta_j(\mathbf{Y},r)=1$, $\delta_j(\mathbf{Y},s)=0$, $\delta_j(\mathbf{Z},r)=0$ and $\delta_j(\mathbf{Z},s)=1$. However, in
both cases (i) and (ii), the product $P^{(j)}(r,s)_{01}P^{(j)}(r,s)_{10}$
appears in the term for $P_j(\mathbf{Y})P_j(\mathbf{Z})$ while
$P^{(j)}(r,s)_{00}P^{(j)}(r,s)_{11}$ appears in the term for $P_j(\mathbf{Y}\vee \mathbf{Z}) P_j(\mathbf{Y}\wedge \mathbf{Z})$. The former term is less than or equal to the
second since
$P^{(j)}(r,s)_{00}P^{(j)}(r,s)_{11}- P^{(j)}(r,s)_{01}P^{(j)}(r,s)_{10} = \det P^{(j)}(r,s)$, which is non-negative
by our assumption. Consequently, all the
terms in $P_j(\mathbf{Y})P_j(\mathbf{Z})$ are less than or equal to the
corresponding terms in $P_j(\mathbf{Y}\vee \mathbf{Z}) P_j(\mathbf{Y}\wedge \mathbf{Z})$. This
establishes (\ref{ineq1}).

We now recall a form of the `four functions theorem', a classical result of Ahlswede and Daykin \cite{ahb}.
Let $(\LL,\vee,\wedge)$ be a finite distributive lattice and let $\alpha$ be a function that assigns a non-negative real number
to each element of $\LL$. For a subset $\A\subseteq\LL$, set $\alpha(\A)=\sum_{A\in\A}\alpha(A)$. If $\alpha$ satisfies the property
that for any two elements $A,B$ of $\LL$, $\alpha(A)\alpha(B) \leq \alpha(A \vee B)\alpha(A \wedge B)$, then
\begin{equation}
\label{4fun}
\alpha(\A)\alpha(\B)\leq\alpha(\A\vee\B)\alpha(\A\wedge\B),
\end{equation}
where $\A\vee\B=\{A\vee B : A\in\A , B\in\B\}$ and $\A\wedge\B=\{A\wedge B : A\in\A, B\in\B\}$.

We apply this theorem by taking $\LL=\LL_V, \alpha=P$ and noting that
$\alpha$ satisfies the required hypothesis by (\ref{ineq1}).
Consider any fixed non-empty subset $W$ of $X$ and recall the definition (for $\mathbf{u}\in\LL_W$) of
$\A_{\mathbf{u}}$ in (\ref{Asets}). Note that:
$$\A_{\mathbf{y}} \vee \A_{\mathbf{z}} = \A_{\mathbf{y}\vee\mathbf{z}}, \mbox{ and } \A_{\mathbf{y}} \wedge \A_{\mathbf{z}} = \A_{\mathbf{y}\wedge\mathbf{z}}.$$
Thus, taking $\A=\A_{\mathbf{y}}$ and $\B = \A_{\mathbf{z}}$ in (\ref{4fun}) we deduce
that:
$$
\alpha(\A_{\mathbf{y}})\alpha(\A_{\mathbf{z}})\leq\alpha(\A_{\mathbf{y}\vee\mathbf{z}})\alpha(\A_{\mathbf{y}\wedge\mathbf{z}}),
$$
which is, by $\alpha=P$ and (\ref{Asets}), equivalent to $p(\mathbf{y})p(\mathbf{z})\leq p(\mathbf{y}\vee\mathbf{z})p(\mathbf{y}\wedge\mathbf{z})$.
\end{proof}

\section{Application: Predicting future phylogenetic diversity}
\label{applic}

\subsection{Expected future phylogenetic diversity}
\label{expect}

In this section, we use Proposition~\ref{subadd} to obtain an inequality concerning the expected loss of biodiversity under species extinction models. Consider a rooted directed tree $T=(V_T,A_T)$ in which all the arcs are directed away from the root and with leaf set $X$. Let each
arc $a$ in $A_T$ be assigned a non-negative length $\lambda_a$. Here, $T$ represents the evolutionary history of the species in $X$, while $\lambda_a$ refers either to the amount of the genetic change on arc $a$, to its temporal duration or to some other feature such as morphological diversity.

Given a subset $Y$ of $X$, the {\em phylogenetic diversity} (PD) of
$Y$, denoted $\varphi_Y$, is the sum of the lengths of the arcs of
the minimal subtree of $T$ that connects the root and the leaves in
$Y$. PD has been widely used to measure the biodiversity of a group of species \cite{fai, fal, nee, stf}; informally, the PD-score of a subset $Y$ measures how much total `genetic' or `evolutionary' diversity in the tree is spanned just by the
the species in $Y$ (depending on whether the lengths assigned to the edges reflect the amount of genetic change or evolutionary time, respectively).

As a function from $2^X$ to $\RR^{\geq0}$ $\varphi$ has some attractive properties for the discrete mathematician: as well as being a submodular, increasing function it also has the property that the subsets of $X$ of given cardinality that have maximal
$\varphi$ value form a (strong) greedoid, and so can be quickly constructed by the greedy algorithm (for details, see \cite{mou}).

\begin{figure}[htp]
\begin{center}
\begin{pspicture}(0,-0.25)(12,2.25)

\dotnode(2.25,2.25){p1}
\dotnode(0.75,1.25){p2}
\dotnode(3,2){p3}
\dotnode(0,0.75){p4}
\dotnode(1.25,0.25){p5}
\dotnode(2,1){p6}
\dotnode(2.25,0.5){p7}
\dotnode(2.5,1.5){p8}
\dotnode(3,0.75){p9}
\dotnode(3.25,1.5){p10}
\dotnode(4,0){p11}
\dotnode(5.25,1.25){p12}

\ncline[linewidth=0.02]{p1}{p4}
\ncline[linewidth=0.02]{p2}{p5}
\ncline[linewidth=0.02]{p1}{p12}
\ncline[linewidth=0.02]{p3}{p11}
\ncline[linewidth=0.02]{p3}{p6}
\ncline[linewidth=0.02]{p7}{p8}
\ncline[linewidth=0.02]{p9}{p10}

\pcline{->}(5.75,1.05)(6.25,1.05)

\dotnode(9,2.25){q1}
\dotnode(7.5,1.25){q2}
\dotnode(9.75,2){q3}
\dotnode(6.75,0.75){q4}
\dotnode(8,0.25){q5}
\dotnode(8.75,1){q6}
\dotnode(9,0.5){q7}
\dotnode(9.25,1.5){q8}
\dotnode(9.75,0.75){q9}
\dotnode(10,1.5){q10}
\dotnode(10.75,0){q11}
\dotnode(12,1.25){q12}

\ncline[linewidth=0.02]{q1}{q2}
\ncline[linewidth=0.01,linestyle=dashed]{q4}{q2}
\ncline[linewidth=0.02]{q2}{q5}
\ncline[linewidth=0.02]{q1}{q3}
\ncline[linewidth=0.01,linestyle=dashed]{q3}{q12}
\ncline[linewidth=0.02]{q3}{q11}
\ncline[linewidth=0.02]{q3}{q8}
\ncline[linewidth=0.01,linestyle=dashed]{q8}{q6}
\ncline[linewidth=0.02]{q7}{q8}
\ncline[linewidth=0.01,linestyle=dashed]{q9}{q10}

\rput(1.25,0){\rnode{a1}{*}}
\rput(2.25,0.25){\rnode{a1}{*}}
\rput(4,-0.25){\rnode{a1}{*}}

\end{pspicture}
\end{center}\caption{If only the species marked * in the tree on the left survive then the future PD is the sum of the lengths of the solid edges in the tree on the right.} \label{peedee}
\end{figure}
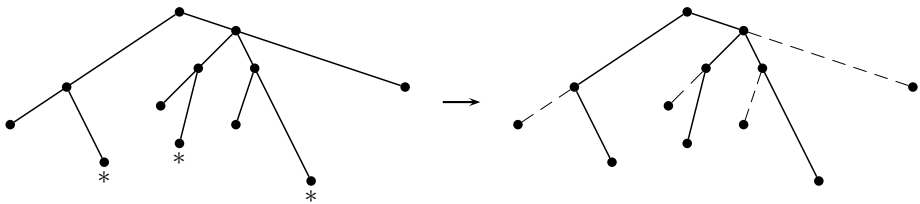

Assume that species in $X$ undergo random extinction and let $E_x$ denote the event that a species
$x\in X$ is extinct at some fixed future time $t$.  Consider the phylogenetic diversity $\varphi$ of the group of species that are still extant at time $t$. This random variable is referred to as {\em future PD}. An example to illustrate this notion is given in Figure~\ref{peedee}.

The expected value of $\varphi$ is:
\begin{equation}
\label{ee}
\EE[\varphi]= \sum_{a=(u,v) \in A_T} \lambda_a\cdot(1-\PP(\bigcap_{x \in C_v}E_x))= \varphi_X-\sum_{a=(u,v) \in A_T} \lambda_a\cdot\PP(\bigcap_{x \in C_v}E_x),
\end{equation}
where $C_v$ denotes the subset of $X$ which is separated from the
root by $v$ and which equals $\{v\}$ if $v$ is a leaf vertex. $\EE[\varphi]$ is referred to as {\em expected future PD}.

In the {\em generalized
field of bullets model} (g-FOB) \cite{fal}, the events $E_x^{(g)}:=E_x$ are
independent, and so the probability
$\PP(\bigcap_{x \in C_v}E_x^{(g)})$ that all the species descended from $v$ become extinct can be written as:
\begin{equation}
\label{peqs}
\PP(\bigcap_{x \in C_v}E_x^{(g)}) =  \prod_{x \in C_v}p_x,
\end{equation}
where $p_x$ denotes the probability $\PP(E_x^{(g)})$.

However, the assumption that the events $E_x$ are independent is likely to be
unrealistic in most settings (see, for example, \cite{hea, sim}). In particular, rates at which lineages become extinct may be influenced by some species traits \cite{mad, fitz}.
The model referred to as the {\em state-based field of bullets model} (s-FOB) in \cite{stf} is based on the idea that closely related species in $T$ are more
likely to share attributes that may put them at risk in a hostile
future environment. It assumes that the extinction risk of each species is influenced by some associated binary state with values $0$ and $1$, where state $0$ confers an elevated risk of extinction for example under climate change.

Here, we generalize this model and suppose that the extinction risk of each species
$x$ is influenced by $k$ binary states, each of which takes values in $\{0,1\}$, where state $1$ is always advantageous over state $0$ for $x$. We suppose that it is not known what features will help species survive and so the states are not known for the species in $X$.  However, we assume that the $k$ states have evolved
under $k$ independent Markovian models on $T$ assigning a state in $\{0,1\}^k$ to each species.

We assume further that if the states were determined at
the leaves, then extinction would proceed according to the g-FOB model
in which species $x$ is extinct at time $t$ with probability $p_x^\mathbf{i}$ if
it is in state $\mathbf{i} \in \{0,1\}^k$. Finally, we suppose that for each species $x\in X$ and any two states $\mathbf{i}=(i_1,\ldots,i_k)$ and $\mathbf{l}=(l_1,\ldots,l_k)$:
\begin{equation}
\label{kfob}
p_x^{\mathbf{i}}\leq p_x^{\mathbf{l}} \mbox{   whenever   } l_j\leq i_j \mbox{   for each   } j=1,\ldots,k.
\end{equation}
This condition says that state $\mathbf{l}$ confers at least as high an extinction risk on a species $x$ as state $\mathbf{i}$ if all the binary states in $\mathbf{i}$ are at least as `advantageous' for $x$ as the binary states in $\mathbf{l}$. Note, however, that if condition $l_j\leq i_j$ is not satisfied for every $j$, there is no prescribed relationship between $p_x^{\mathbf{i}}$ and $p_x^{\mathbf{l}}$. We have the freedom to specify these relationships according to the needs of the model being studied, or leave them unspecified. For example, we may assume that the $k$ binary states are ordered in a decreasing manner by their importance for survival and that $p_x^{\mathbf{i}}\leq p_x^{\mathbf{l}}$, whenever $l_j\leq i_j$ for the smallest coordinate $j\in\{1,\ldots,k\}$ for which $i_j\ne l_j$. Alternatively, we may assume that all the states are equally important for survival and that $p_x^{\mathbf{i}}\leq p_x^{\mathbf{l}}$, whenever $\sum_{j=1}^k l_j\leq \sum_{j=1}^k i_j$; that is, the more coordinates of the state assigned to $x$ are $1$ the smaller is the extinction probability of $x$. In the following, we only assume the relationships described in (\ref{kfob}).

We call the model described above the {\em trait-dependent
field of bullets} model (t-FOB). In the case when $k=1$, this model is the s-FOB model, whereas the case where for each $x$, $p_x^{\mathbf{i}}=p_x^{\mathbf{l}}$ for any two states $\mathbf{i},\mathbf{l}\in\{0,1\}^k$ gives the g-FOB model.

Given a t-FOB model, consider the g-FOB model in which
the extinction probability of each species $x$ is the same as in the
t-FOB model. That is, if {\boldmath $\xi$} describes the
multivariate Markov process and the values $p_x^{\mathbf{i}}$ are the conditional extinction probabilities in the t-FOB model, then, in the associated g-FOB model, each species $x\in X$ goes extinct with probability
\begin{equation}
\label{px}
p_x =\PP[E_x^{(g)}]=\PP[E_x^{(t)}] =\sum_{\mathbf{i}\in\{0,1\}^k} p_x^{\mathbf{i}}\PP(\mbox{\boldmath $\xi$(x)}=\mathbf{i}),
\end{equation}
where $E_x^{(t)}$ denotes the event $E_x$ under t-FOB. Theorem~\ref{loss} compares the loss of PD under a t-FOB model with the PD loss under the associated g-FOB model.

\begin{theorem}
\label{loss}
Consider a t-FOB model on a fixed tree $T$ with non-negative arc lengths and with leaf set $X$.
The expected future
PD of this model is less than or equal to the expected future PD of the
associated g-FOB model. 
\end{theorem}

\begin{proof}
Let {\boldmath $\xi$} and $p_x^{\mathbf{i}}$ denote the Markov process and the extinction probabilities of the t-FOB model, respectively.
In view of (\ref{ee}) and (\ref{peqs}), it suffices to show that:
\begin{equation}
\label{ine1}
\prod_{x \in C_v} p_x \leq \PP(\bigcap_{x \in C_v}E_x^{(t)}) ,
\end{equation}
where
$p_x$ is given in (\ref{px}).
Recall how we defined the lattice $(\LL_W,\vee,\wedge)$ for a Markov process on a tree and for a non-empty subset $W$ of the leaf set of the tree in the previous section, and consider $(\LL_{C_v},\vee,\wedge)$. 
Since, for $\mathbf{u}\in \LL_{C_v}$, $p(\mathbf{u})$ denotes the probability that for each $x\in C_v$, $x$ is assigned $\mathbf{u}(x)\in\{0,1\}^k$, we get:
$$\PP(\bigcap_{x \in C_v}E_x^{(t)}) = \sum_{\mathbf{u} \in \LL_{C_v}} p(\mathbf{u})\prod_{x
\in C_v}f_x(\mathbf{u}),$$
where $f_x(\mathbf{u})$ is the probability that $x$ becomes extinct given that it is in state $\mathbf{u}(x)$; that is, $f_x(\mathbf{u})=p_x^{\mathbf{u}(x)}$.
Moreover, for each $x\in C_v$, we have:
\begin{eqnarray}
p_x = \sum_{\mathbf{i}\in\{0,1\}^k} p_x^{\mathbf{i}}\PP(\mbox{\boldmath $\xi$}(x)=\mathbf{i})
= \sum_{\mathbf{i}\in\{0,1\}^k} p_x^{\mathbf{i}}\left(\sum_{\mathbf{u}\in \LL_{C_v}:\mathbf{u}(x)=\mathbf{i}} p(\mathbf{u})\right)
= \sum_{\mathbf{u}\in \LL_{C_v}} p(\mathbf{u}) f_x(\mathbf{u}).\nonumber
\end{eqnarray}
Now we can rewrite (\ref{ine1}) as
\begin{equation}
\label{ine2}
\prod_{x \in C_v} \left( \sum_{\mathbf{u}\in \LL_{C_v}} p(\mathbf{u}) f_x(\mathbf{u})\right) \leq \sum_{\mathbf{u} \in \LL_{C_v}} p(\mathbf{u})\prod_{x \in C_v}f_x(\mathbf{u}).
\end{equation}
The proof of (\ref{ine2}) makes use of
Proposition~\ref{subadd} as well as the following multivariate form of the FKG inequality of Fortuin,
Kasteleyn and Ginibre (1971) \cite{fkg}.
Given a finite distributive lattice $(\LL,\vee,\wedge)$, suppose that $f_1,f_2,\ldots,f_n$ are
functions from $\LL$ into the non-negative real
numbers that satisfy, for any two elements $A$ and $B$ of $\LL$, the condition that:
\begin{equation}
\label{inclusion}
A \leq B \Rightarrow f_i(A) \geq f_i(B).
\end{equation}
Furthermore, suppose that $\mu$ is a probability measure on the elements of $\LL$ which satisfies the condition that
\begin{equation}
\label{inclusion2}
\mu(A)\mu(B) \leq \mu(A \vee B)\mu(A \wedge B)\mbox{  for any pair  } A,B\in \LL.
\end{equation}
Then:
\begin{equation}
\label{ine3}
\prod_{i=1}^n\left(\sum_{A\in\LL}\mu(A)f_i(A)\right) \leq
\sum_{A\in\LL} \mu(A)\prod_{i=1}^n f_i(A).
\end{equation}

We apply this inequality  by
setting $\LL=\LL_{C_v}$, $\mu = p$ and $f_x(\mathbf{u})=p_x^{\mathbf{u}(x)}$ for $\mathbf{u}\in\LL_{C_v}$, $x\in C_v$.
Note that $f_x$ satisfies (\ref{inclusion}). Namely, $\mathbf{u}\leq\mathbf{y}$ (for $\,\mathbf{u},\mathbf{y}\in\LL_{C_v}$) means that $u_j(x)\leq y_j(x)$ for each coordinate $j$, which, by (\ref{kfob}), implies $p_x^{\mathbf{u}(x)}\geq p_x^{\mathbf{y}(x)}$.
Note also that $\mu$ satisfies (\ref{inclusion2}) by Proposition 2.1.
In view of these, (\ref{ine3}) provides inequality (\ref{ine2}), and the proof is complete.
\end{proof}

\subsection{Variance of future PD}
\label{var}

Consider now the variance of $\varphi$:
\begin{equation}
\label{genvar}
\V[\varphi]=\Cov[\varphi,\varphi]=\sum_{a,b\in A_T}\lambda_a\lambda_b \Cov[Y_a,Y_b],
\end{equation}
where $Y_a$ is the random variable that takes value $1$ if arc $a$ is part of the subtree connecting the survival species and the root and takes value $0$ otherwise. Our goal is to compare the variance under a t-FOB model to the variance under the associated g-FOB model. It is easy to find examples in which the former variance is greater than the latter and so we will only show 
that the variance for a t-FOB model can be less than that of the associated g-FOB model. To this end, let $T$ be the tree with leaf set $\{x,y\}$ in which the arcs $b$ and $c$ pointing to $x$ and $y$, respectively, are incident with the single interior vertex of the tree, which is adjacent to the root by arc $a$.
Consider $\Cov[Y_a,Y_a]=(1-\PP[E_x\cap E_y])\PP[E_x\cap E_y]$, which is written as $(1-\PP[E_x^{(t)}\cap E_y^{(t)}])\PP[E_x^{(t)}\cap E_y^{(t)}]$ in t-FOB and which becomes $(1-p_xp_y)p_xp_y$ under g-FOB. Note that $\Cov[Y_a,Y_a]$ is less under a t-FOB than under the associated g-FOB if and only if $\PP[E_x^{(t)}\cap E_y^{(t)}]>p_xp_y$ and $\PP[E_x^{(t)}\cap E_y^{(t)}]+p_xp_y>1$ hold. It is easy to see that these conditions can be satisfied by some t-FOB model (together with its g-FOB) on $T$. Additionally, for any such t-FOB model, a value of $\lambda_a$ can be chosen that is large enough in relation to $\lambda_b$ and $\lambda_c$ so that $\lambda_a^2 \Cov[Y_a,Y_a]$ is the dominant term in (\ref{genvar}), resulting in a greater total variance for the corresponding g-FOB.

The following example describes an s-FOB (that is, a t-FOB with $k=1$) under which the variance is less than the variance under the associated g-FOB.

\begin{figure}[htp]
\begin{center}
\begin{pspicture}(0,0)(2,4.5)

\dotnode(0.25,0.25){s1}
\dotnode(1.75,0.25){s2}
\dotnode(1,1){s3}
\dotnode(1,4){s4}

\ncline[linewidth=0.02]{s1}{s3}
\ncline[linewidth=0.02]{s2}{s3}
\ncline[linewidth=0.02]{s3}{s4}

\rput(0.25,0){\rnode{m1}{$x$}}
\rput(1.75,0){\rnode{m2}{$y$}}
\rput(1.15,2.5){\rnode{m3}{$a$}}
\rput(0.5,0.75){\rnode{m4}{$b$}}
\rput(1.5,0.75){\rnode{m5}{$c$}}
\rput(1,4.25){\rnode{m6}{root}}

\end{pspicture}
\end{center}\caption{\label{varexample}}
\end{figure}

\noindent{\bf Example.} Let $T$ be the tree shown in Figure~\ref{varexample} with arc lenghts $\lambda_a=4$ and $\lambda_b=\lambda_c=1$ and consider the following s-FOB model on $T$. Let $\xi$ be a two-state Markov process on $T$ with the state space $\{0,1\}$ so that $\pi_0=\pi_1=\frac{1}{2}$ and each arc is assigned the transition matrix
$\bigl(\begin{smallmatrix} \frac{3}{4} & \frac{1}{4} \\ \frac{1}{4} & \frac{3}{4}
\end{smallmatrix}\bigr)$.
Let $p_x^0=p_y^0=\frac{7}{8}$ and $p_x^1=p_y^1=\frac{6}{8}$. A careful check shows that the variance under this model is less than the variance under the associated g-FOB model (in which $p_x=p_y=\frac{13}{16}$).

\end{document}